\documentclass[sigconf,nonacm]{acmart}

\usepackage{graphicx}
\usepackage{algorithm}
\usepackage[noend]{algpseudocode}
\usepackage{amsmath}
\usepackage{multirow}
\usepackage{graphicx}
\usepackage{xcolor}
\usepackage{amsfonts}
\usepackage{amsthm}
\usepackage{pifont}
\usepackage{xspace}
\usepackage{mathtools}
\usepackage{enumitem}
\usepackage{lineno}
\usepackage{thmtools}
\usepackage{thm-restate}
\usepackage{relsize}
\usepackage{hhline}
\usepackage{etoolbox}
\usepackage{thmtools}
\usepackage{thm-restate}
\usepackage{cleveref}
\usepackage[most]{tcolorbox}
\usepackage{multicol}
\usepackage{changepage}
\usepackage{algpseudocode}
\usepackage{hhline}
\usepackage{etoolbox}
\usepackage{relsize}
\usepackage{bm}
\usepackage{scalefnt}
\nolinenumbers 

\setlist[description]{leftmargin=*,labelindent=*}

\newcommand{\alglinenoNew}[1]{\newcounter{ALG@line@#1}}
\newcommand{\alglinenoPop}[1]{\setcounter{ALG@line}{\value{ALG@line@#1}}}
\newcommand{\alglinenoPush}[1]{\setcounter{ALG@line@#1}{\value{ALG@line}}}

\newtheorem{observation}{Observation}
\newtheorem{proposition}{Proposition}

\newcommand{\mypara}[1]{\smallskip\noindent\textbf{#1.}}

\newcommand{\ia}{\textit{i}}
\newcommand{\ib}{\textit{ii}}
\newcommand{\ic}{\textit{iii}}
\newcommand{\id}{\textit{iv}}

\newcommand{\com}[1]{}

\algdef{SE}[Receiving]{Receiving}{EndReceiving}[1]{\textbf{upon
		receiving}\ #1\ \algorithmicdo}{\algorithmicend\ \textbf{}}%
\algtext*{EndReceiving}

\algdef{SE}[Upon]{Upon}{EndUpon}[1]{\textbf{upon}\ #1\ \algorithmicdo}{\algorithmicend\ \textbf{}}%
\algtext*{EndUpon}

\newcommand\StateX{\Statex\hspace{\algorithmicindent}}
\newcommand\StateXX{\StateX\hspace{\algorithmicindent}}
\algrenewcommand\textproc{}

\newcommand{\oded}[1]{\textcolor{red}{[Oded: #1]}}
\newcommand{\udi}[1]{\textcolor{blue}{[Udi: #1]}}

\newcommand{\andy}[1]{\textcolor{purple}{[Andy: #1]}}

\newcommand{\temph}[1]{\emph{#1}}

\makeatletter \let\sv@thm\@thm \def\@thm{\let\indent\relax\sv@thm} \makeatother

\newcommand{\calS}{\mathcal{S}}

\newcommand{\calB}{\mathcal{B}}

\newcommand{\calX}{\mathcal{X}}

\crefname{table}{table}{tables}
\crefname{table}{Table}{Tables}
\crefname{algocf}{alg.}{algs.}
\crefname{algocf}{Alg.}{Algs.}
\crefname{figure}{Fig.}{Figs.}
\crefname{figure}{fig.}{figs.}
\crefname{claim}{claim}{claims}
\crefname{claim}{Claim}{Claims}
\crefformat{chapter}{\S#2#1#3}
\crefmultiformat{chapter}{\S\S#2#1#3}{and~#2#1#3}{, #2#1#3}{, and~#2#1#3}

\crefformat{section}{\S#2#1#3}
\crefmultiformat{section}{\S\S#2#1#3}{and~#2#1#3}{, #2#1#3}{, and~#2#1#3}

\usepackage{enumitem}
\setlist{nosep} 
\setlist{itemsep=1pt, topsep=3pt}

\newcommand{\sys}{Flash\xspace}

\AtBeginDocument{%
  }

\begin{document}

\title[Flash: An Asynchronous Payment System]{Flash: An Asynchronous Payment System \\ with Good-Case Linear Communication Complexity}

\author{Andrew Lewis-Pye}
\affiliation{%
 \institution{London School of Economics}
   \country{United Kingdom}
 }

\author{Oded Naor}
\affiliation{%
 \institution{Technion and StarkWare}
  \country{Israel}
}

\author{Ehud Shapiro}
\affiliation{%
 \institution{Weizmann Institute of Science}
   \country{Israel}
}

%
%
%
%
%
%
%

\renewcommand{\shortauthors}{Lewis-Pye, Naor, Shapiro}

\settopmatter{printacmref=false}

\begin{abstract}
While the original purpose of blockchains was to realize a payment system, it has been shown that, in fact, such systems do not require consensus and can be implemented deterministically in asynchronous networks.   State-of-the-art payment systems employ Reliable Broadcast to disseminate payments and prevent double spending, which entails $O(n^2)$ communication complexity per payment even if Byzantine behavior is scarce or non-existent.

Here we present \emph{\sys}, the first payment system to achieve $O(n)$ communication complexity per payment in the good case and $O(n^2)$ complexity in the worst-case, matching the lower bound.  This is made possible by sidestepping Reliable Broadcast and instead using the blocklace---a DAG-like partially-ordered generalization of the blockchain---for the tasks of recording transaction dependencies, block dissemination, and equivocation exclusion, which in turn prevents doublespending. 

\sys has two variants: for high congestion when multiple blocks that contain multiple payments are issued concurrently; and for low congestion when payments are infrequent.
\end{abstract}

\received{20 February 2007}
\received[revised]{12 March 2009}
\received[accepted]{5 June 2009}

\maketitle

\section{Introduction}
The rise of \emph{cryptocurrencies} and \emph{blockchains} such as Bitcoin~\cite{bitcoin}, Ethereum~\cite{buterin2014next}, and more, has changed the way people can use financial services and interact among themselves.
To date, Bitcoin is the most popular blockchain~\cite{coingecko} and is still used primarily for money transfers.

A \emph{payment system} or \emph{asset transfer} is an abstraction first formally defined by Guerraoui et al. in 2019~\cite{guerraoui2019consensus}.
Its goal is to capture the core motivation behind Bitcoin as a payment system.
In this abstraction, there is a set of agents, each associated with an account with an initial balance, and each agent can transfer money from its own account to other accounts.
The abstraction also allows any agent to check the balance of any account, similar to how a blockchain is public, and any entity with access to it can read its entire data.

While almost all cryptocurrencies implement some form of \emph{consensus} to totally order the transactions in the blockchain, Guerraoui et al.~\cite{guerraoui2019consensus} prove that a payment system is a weaker problem, i.e., payments with no dependencies between them do not have to be ordered.
They also provide a concrete implementation of a payment system in an asynchronous network, circumventing the seminal FLP result~\cite{fischer1985impossibility}, which proves that consensus cannot be implemented deterministically in asynchrony.

Existing algorithms for payment systems, including the ones in~\cite{guerraoui2019consensus,auvolat2020money,collins2020online}, are based on \emph{reliable broadcast (RB)}~\cite{bracha1987asynchronous} to disseminate blocks to other agents.
RB is an abstraction that prevents Byzantine agents from equivocating, which in the context of a payment system prevents Byzantine agents from doublespending.

If $n$ is the number of agents, then existing solutions have an $O(n^2)$ communication complexity per payment, which matches the known lower bound~\cite{naor2022payment}.
This is also the communication complexity of these protocols even in case the number of Byzantine agents is low, or even if there are no such parties at all.
In real systems, the worst case rarely happens, and therefore algorithms that are safe even when facing the worst case, but optimize for performance in the good case, are highly desirable.

Within this context, we present \emph{\sys}, a payment system that optimizes the communication complexity for the good case and matches the known lower bound in the worst case. In the following, we let $f$ be a known bound on the number of Byzantine agents, and let $t\leq f$ be the (unknown) actual number of Byzantine agents.
\sys always works in asynchrony, but in the good case where $t$ is constant (or zero, i.e., no malicious behavior at all) and the communication is synchronous, then \sys has an $O(n)$ complexity.
In the worst case, where $t \in \Theta(n)$, the complexity is $O(n^2)$.
This matches the lower bound~\cite{naor2022payment} in which a single payment requires $O(n^2)$ complexity in the worst-case.

To achieve this performance, \sys forgoes the use of RB to disseminate blocks and to exclude equivocations along the way, and uses instead the \emph{blocklace}.
The blocklace is a distributed DAG-like structure built jointly by the agents.
The vertices in the blocklace are blocks containing payments, and the edges are hash pointers from each block to previous blocks in the blocklace.

The blocklace is used for all tasks associated with building a payment system.
It is used to disseminate blocks (that might contain equivocations since RB is not used), represent the relations and dependencies between payments, and lastly, to exclude equivocations that might otherwise cause doublespends from the same account.

\begin{figure}[!ht]
\centering
\includegraphics[width=8cm]{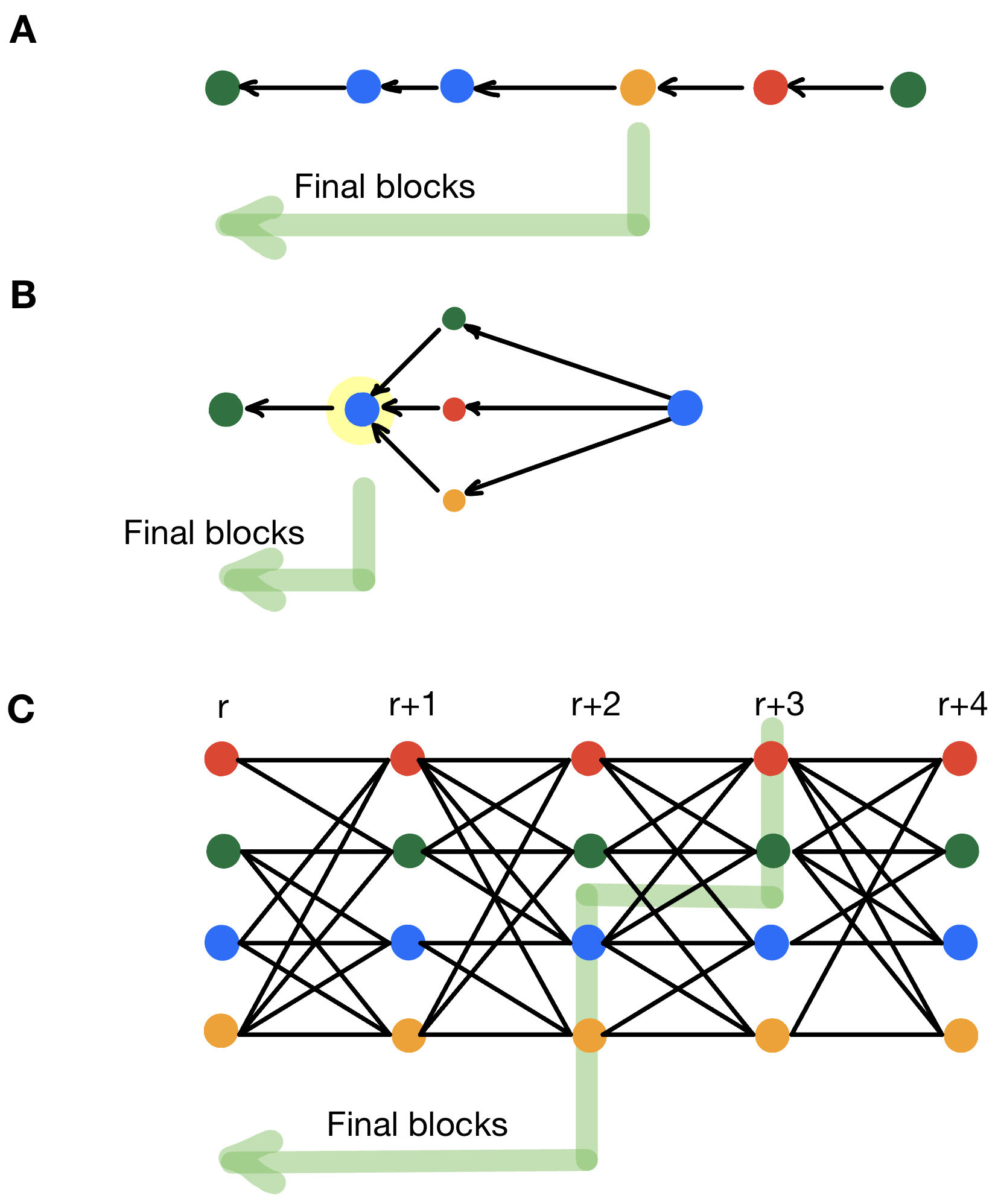}
\caption{The Flash payment system with four agents, at most one of which is Byzantine. Good-case runs.  Transactions in ratified blocks are final. A. Low-Congestion Protocol.  B. Example of an urgent block (with a yellow halo).  C.  High-Congestion Protocol. 
}

\label{figure:goodcase}
\end{figure}

We present two variants of \sys.
The first is designed for the case where agents issue payments infrequently, so that the expected gap between payments (say tens of seconds) is significantly larger than network latency (say tens of ms). An example scenario would be a  digital community bank running on its members' smartphones, in which the number of transactions is in the thousands per day.
In this variant,  an agent that wishes to issue a payment  simply creates a new block and sends it to all other agents. 
Assuming that collisions between concurrent payments are rare, the backbone of the blocklace  is similar to a single linear blockchain (each block typically with additional input dependencies on earlier blocks , see Fig. \ref{figure:goodcase}.A). 

Each block in this blocklace, in addition to encoding a transaction, is also used to  approve  previous blocks.
Thus, each transaction is final  once it has $\frac{1}{2}(n+f)$ succeeding payments from different agents that approve it.
The downside of this approach is that, if indeed transactions are infrequent, then it can take a long time to finalize a single transaction.

To solve this, we give an option of \emph{urgent payments}, in which an agent can ``mark'' a limited number of transactions as urgent, and then all honest agents approve it upon receipt, without waiting until they have to issue a payment on their own.  Equipped with this option, the blocklace structure can be depicted as seen in Fig. \ref{figure:goodcase}.B.

The second variant of \sys is for a scenario where payment frequency is high and concurrent payments are the norm.
For example, when the agents are banks issuing payments to each other on behalf of their customers, as in the SWIFT network.
In this variant, a block is created only after its creator has observed new blocks from $(n+f)/2$ of the agents, and the blocklace looks similar to the one depicted in Fig. \ref{figure:goodcase}.C.
Using this variant, each transaction may contain many payments by the same agent (e.g. a bank issuing payments to other banks on behalf of its customers, see above), and many transactions may be issued concurrently, supporting
an overall high frequency of payments.

The resulting blocklace structure in this variant is similar to that of the Cordial Miners consensus protocol~\cite{keidar2022cordial}, and is effectively divided into rounds, where each round contains $O(n)$ blocks. 
The main difference is that, since ordering consensus is not needed  to implement a payment system,  many of the complexities of Cordial Miners can be stripped away.

In this variant, in the good case it takes between one and two rounds of the blocklace in expectation to finalize a payment issued by an honest agent. 
By batching in each block a linear number of payments, the protocol achieves an amortized $O(n)$ communication complexity per single payment and fast finality. 

\mypara{Roadmap} The rest of the paper is structured as follows: \Cref{sec:model} formally defines the model and the payment system asbtraction; \Cref{sec:transactionBlocklace} describes the transaction blocklace data structure; \Cref{sec:Flash} presents the \sys payment system algorithms for low congestion (\Cref{sec:Flash:lowCongestion}) and high congestion (\Cref{sec:Flash:highCongestion}); \Cref{sec:relatedWork} is related work; and  \Cref{sec:conclusion} concludes the paper.
Blocklace utilities are provided as   \Cref{appendix:blocklace}.
\section{Model and Problem Definition} \label{sec:model}
The payment system consists of a set of agents $P$, $|P| = n$, connected by an asynchronous reliable network,
each of which is endowed with a single and unique key-pair and identified by its public key $p \in P$. 
Up to $f<n/3$ of the agents may be Byzantine.  We adopt the UTXO model of Bitcoin. Each agent $p\in P$ thus begins with an \emph{initial} ``unspent transaction output''  (UTXO) that only $p$ can spend, and which specifies the initial balance of agent $p$.  A transaction signed by an agent $p$ has UTXOs assigned to $p$ as input -- these are the UTXOs that the transaction `spends' --  and UTXOs that $p$ assigns to other agents, including $p$ itself, as output, where the inputs and outputs total to the the same amount.
Two transactions by the same agent that spend a common UTXO are said to constitute a doublespend. Correct agents do not doublespend.

We consider a notion of \emph{finality} for transactions,  which must satisfy the condition that if any correct agent regards a transaction as final, then all correct agents eventually regard it as final. For a given set of final transactions, an agent's \emph{balance} is given by the set of UTXO which are: (i) Either initial (as specified above) or outputs of final transactions, and; (ii) Are unspent. To implement a payment system, the requirements are then:

\noindent \textbf{Safety:} Two transactions that constitute a doublespend are never both final. 

\noindent \textbf{Liveness:} Every transaction produced by a correct agent is eventually final.

\section{Transactions Blocklace} \label{sec:transactionBlocklace}
The Flash Payment System employs the blocklace to realize transactions using notions specified by Alg. \ref{alg:blocklace} (\Cref{appendix:blocklace}) and presented concisely herein.  

A \emph{payment} is a pair $(q,x)$ of amount $x\ge 0$ to $q\in P$.
A \emph{transaction block} $b$ created by an agent $p$, also referred to as a \emph{$p$-block}, \emph{block} or \emph{transaction}, has zero or more payments, referred to as \emph{output payments}, and
zero or more cryptographic hash pointers (\emph{pointers} for short) to other blocks.  
If $b$ has no payments then it is an \emph{ack block}, and
if it has zero  pointers then it is \emph{initial}, in which case it includes a single payment to $p$, referred to as \emph{self-payment}, representing $p$'s initial balance (which could be 0).   A pointer in $b$ may be marked as an \emph{input pointer}, in which case the block it points to is referred as an \emph{input} of $b$ and must include an output payment to $p$. Thus, the set of input pointers is a subset of the set of all pointers.

The $p$-block $b$ is \emph{balanced} if the sum of the amounts of its output payments is equal to the sum of the amounts of payments to $p$ in its inputs.  

Two $p$-blocks constitute a \emph{doublespend} by $p$ if they share an input.  Doublespending is the key issue Byzantine fault  payment systems have to address.

Note that input pointers record transaction dependencies; following them starting from a given transaction leads to all the transactions on which the given transaction depends, and this can be used to verify the internal validity of the transaction.  Non-input pointers are needed for dissemination, equivocation exclusion, and finalizing payments, as explained below.

\begin{figure}[!ht]
\centering
\includegraphics[width=8cm]{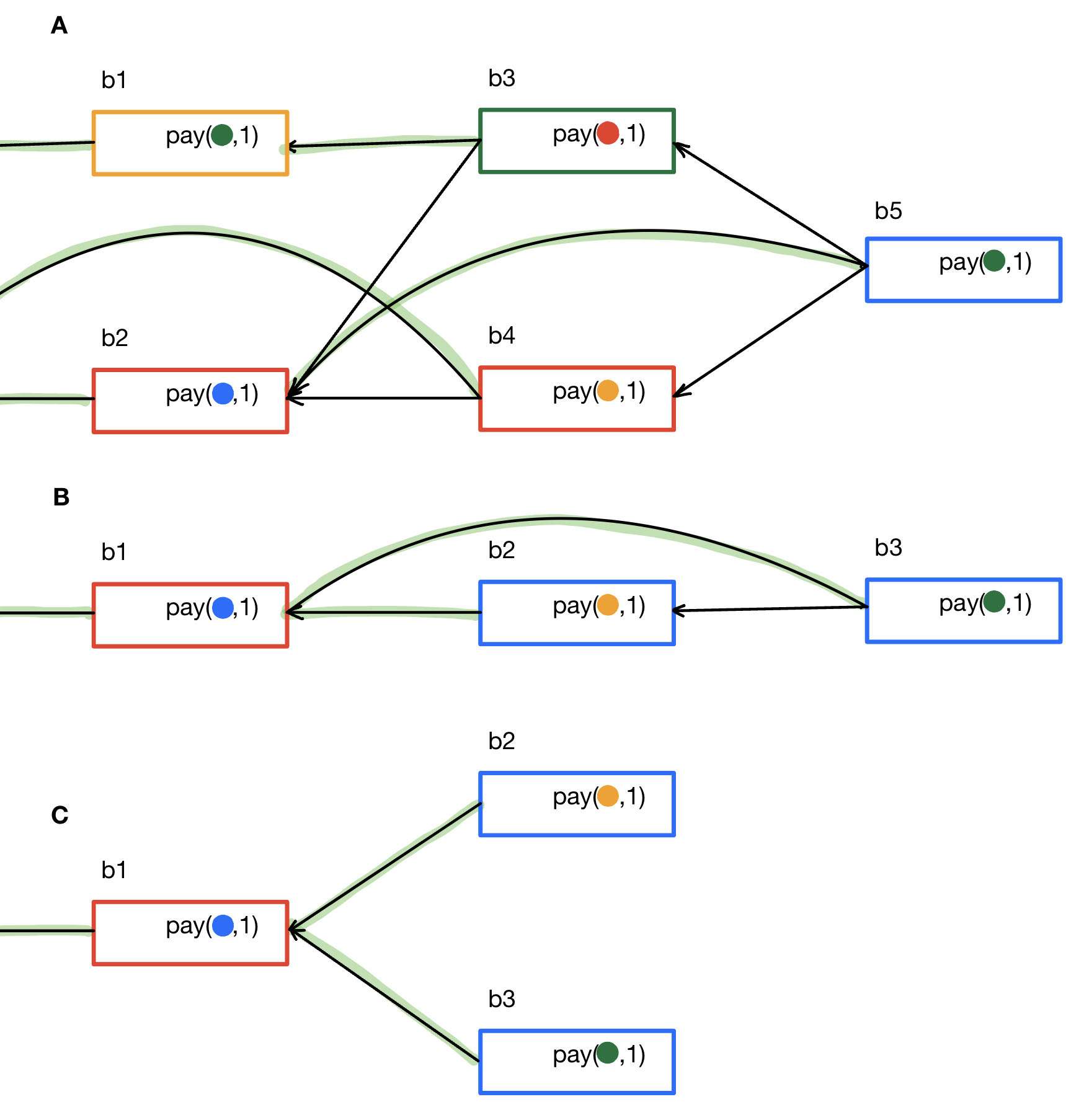}
\caption{Blocks, Transactions, inputs and payments. 
Here, blocks have a single input (marked green) and a single  output payment to a color-coded agent of amount 1.
A. Block $b3$ has pointers to $b1$ and $b2$ but only $b1$ is an input.  Block $b4$ has a pointer to $b2$ and an input pointer to some predecessor of $b2$. 
Block $b5$ has pointers to $b3$ and $b4$ and input pointer to $b2$, which is pointed to by both $b3$ and $b4$.
B. Blocks $b2, b3$ constitute a naive doublespend by the blue agent block $b1$, as $b3$ observes $b2$.
C. Blocks $b2, b3$ constitute a doublespend of the payment in block $b1$ by the equivocation $b2, b3$ of the blue agent.
}
\label{figure:payments}
\end{figure}
A \emph{transactions blocklace}, or \emph{blocklace} for short,  is a set of blocks (See Fig. \ref{figure:payments}.A).
The following notions relate to a given blocklace $B$.  
A block is a \emph{root} of a blocklace if no other block in the blocklace points to it. 
A \emph{path} in $B$ from $b$ to $b'$ is a sequence of blocks in $B$, starting with $b$ and ending with $b'$, each with a pointer to the next block in the path (which we refer to as \emph{predecessor}, according to their order of creation), if any. 
A block $b$  \emph{observes} a block $b'$ in $B$ if there is a path from $b$ to $b'$ in $B$.
A block $b$ \emph{depends} on a block $b'$ if if there is a path with input pointers only from $b$ to $b'$ in $B$ (See Fig. \ref{figure:payments}.A).

 The \emph{closure} of a $p$-block $b$, denoted $[b]$, is the set of blocks it observes; its \emph{input-closure}, denoted by $[b]_i$, is the set of blocks it depends on as well as all the $p$-blocks it observes. 
A blocklace is \emph{closed} if it includes the closure of each of its blocks.
A blocklace $B$ is \emph{input-closed} if it includes the input-closure of each of its blocks.

Two blocks that do not observe each other \emph{equivocate} if by the same agent $p$, in which case $p$ is an \emph{equivocator}, and \emph{collide} if by different agents.  Correct agents do not equivocate.

A $p$-block $b$ is \emph{correct} if it is balanced, $[b]$ includes an initial $p$-block and does not include an equivocation or doublespend by $p$, and every block in $[b]_i\setminus \{b\}$ is correct. Namely a correct block depends only on correct blocks. Correct agents produce only correct blocks.

A block $b$ \emph{approves} $b'$  if it observes $b'$ and does not observe a block equivocating with $b'$.  An agent $p$ \emph{approves} a block $b$ if there is a $p$-block in the blocklace that approves $b$. 

\begin{figure}[!ht]
\centering
\includegraphics[width=7cm]{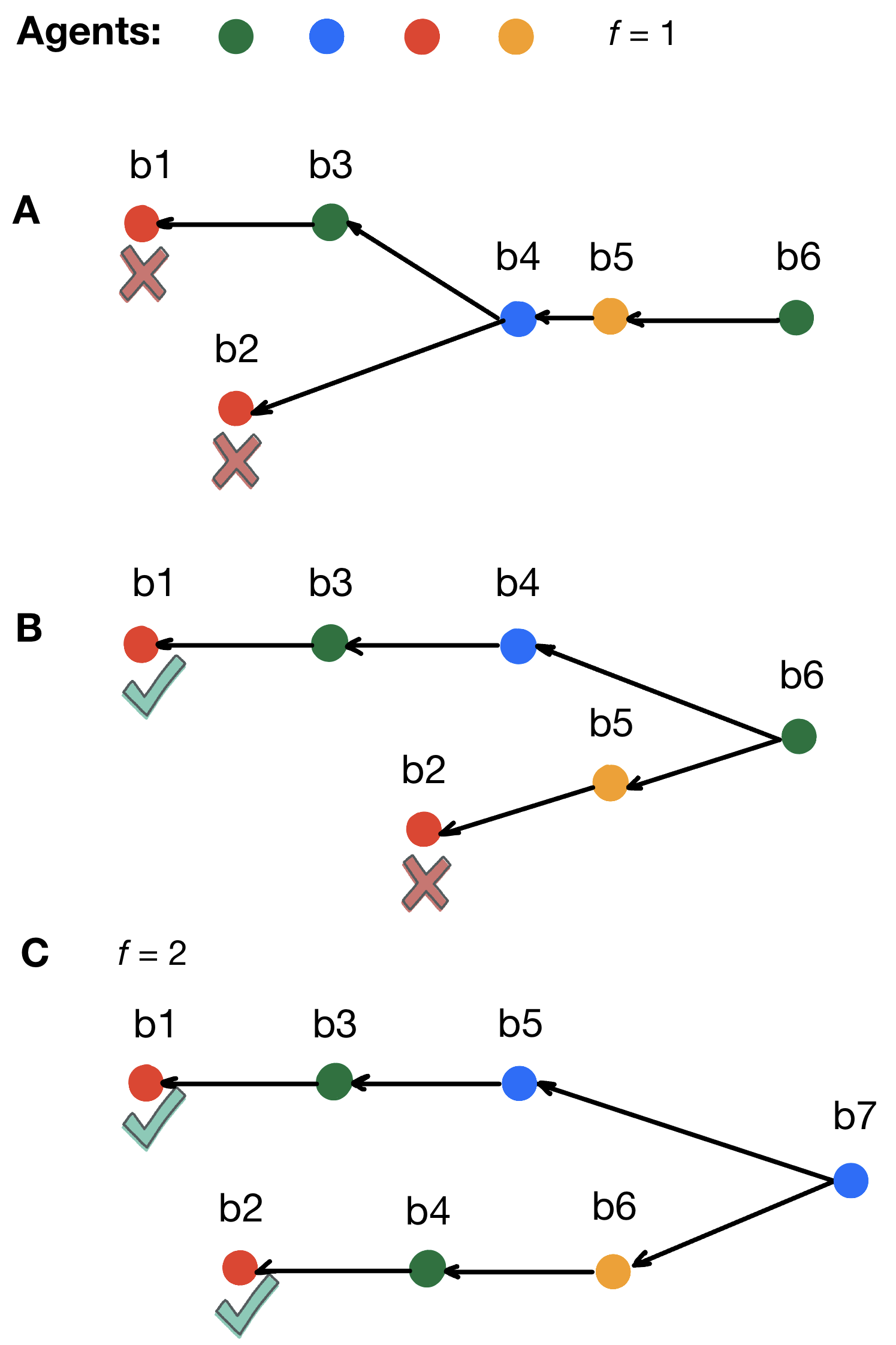}
\caption{Equivocations and their Approval and Finalization. 
Four agents (red, green, blue, yellow) with only the red being faulty ($f=1$).
A. $b1$ and $b2$ are an equivocation by the red agent.  Only the green agent  approves $b1$ via $b3$, which is not a supermajority, as the blue agent ($b4$) immediately sees both $b1$ and $b2$ and thus approves neither.  Hence neither $b1$ nor $b2$ are final.
B.  Both the green ($b3$) and blue ($b4$) agents approve $b1$, which together with the red agent form a supermajority and hence $b1$ is final. $b2$ is approved only by the yellow agent ($b5$), as the green block $b6$ observes both $b1$ and $b2$, and hence $b2$ is not final, so that not both equivocating blocks $b1, b2$ are final. C.  If one more agent equivocates ($f=2$), then the red equivocation $b1, b2$ can become final.
The green agent approves $b1$ with $b3$ and $b2$ with $b4$, equivocating to do so. The blue agent approves $b1$ with $b5$ joining the supermajority (red, green, blue) that finalizes $b1$, and the yellow agent approves $b2$ with $b6$, joining the supermajority (red, green, yellow) that finalizes $b2$. The blue agent then observes  ($b7$) the equivocations of the red and green agents, but it's too late as it has already helped finalize $b1$.
}

\label{figure:approval}
\end{figure}

Note that an agent $p$ may have a $p$-block that approves a block $b$ and a subsequent $p$-block that does not approve $b$ since it observes a block $b'$ that equivocates with $b$.  In this case, the agent $p$ approves $b$ as, according to the definition, it is enough that one $p$-block approves $b$ for $p$ to approve $b$.
For example, consider the green agent in Fig. \ref{figure:approval}.A\&B. 
It has block $b3$ that approves the red block $b1$, and a subsequent block $b6$ that does not approve $b1$ since it observes $b3$  and hence $b1$, but also observes the red block $b2$ equivocating with $b1$, which $b3$ does not observe.  By definition, in such a case the green agent does approve $b1$ (via the block $b2$) but does not approve $b2$, since $b6$ also observes $b1$.

Note that approval is monotonic wrt $\supset$, and that if two blocks equivocate then only an equivocator can approve both of them.
For example, the green agent in Fig. \ref{figure:approval}.C approves the equivocation $b1, b2$ by the equivocation $b3, b4$.

Doublespending and equivocation are related but not identical: An agent may doublespend without equivocating (Fig. \ref{figure:payments}.B), and can equivocate without doublespending (the green agent in Fig. \ref{figure:approval}.C), to maliciously support the doublespending of another agent.  An agent $p$ may doublespend an input payment simply by creating two consecutive (non-equivocating) blocks $b1, b2$ that share the same input  (Fig. \ref{figure:payments}.B); such a doublespend can be promptly detected, as receiving $b2$ requires also receiving its predecessor $b1$.
  The agent $p$ can also doublespend by creating two equivocating blocks $b1, b2$ (Fig.  \ref{figure:payments}.C), sending each to its intended recipient. Detecting and resolving doublespending by equivocation requires collaboration and coordination among the agents, and doing so is the main task of payment systems.

Recall that we assume that at most $f<\frac{1}{3}n$ of the agents may be faulty/Byzantine.
A set of agents forms a \emph{supermajority} if their number is greater than  $\frac{1}{2}(n+f)$.  Note that if $f=0$ then a simple majority is a supermajority, and if $f=\frac{1}{3}(n-1)$ then a supermajority is at least $\frac{2}{3}n$.  
A block $b$ is \emph{final} in a blocklace  $B$ if a supermajority of the agents approve $b$ in $B$ (i.e., there is a supermajority $Q$ of agents such that for each agent $q\in Q$ there exists \emph{some} $q$-block in $B$ that approves $b$).
Note that if there are at most $f$ equivocators in $B$, and if two blocks equivocate in $B$, then it cannot be the case that both blocks are final in $B$ (Proposition \ref{proposition:LC-safety}).  See Fig. \ref{figure:approval}.

Let $B$ be a blocklace.  A payment to $p$ in a block $b\in B$ is an \emph{unspent transaction output} (UTXO), \emph{unspent} for short, if there is no $p$-block in $B$ with $b$ as input.
If $B$ is input-closed and  has only balanced blocks and no doublespends, then the \emph{balance} of an agent $p \in P$ in $B$ is the sum of all unspent payments to $p$ in final blocks in $B$. 


\section{The \sys Payment System} \label{sec:Flash}

Correct agents maintain a closed blocklace, create correct blocks, do not doublespend or equivocate, and disseminate blocks they know to agents that need them.  The difference between the following protocols is in the decision when to issue a block and what to include in it.

\subsection{The Low-Congestion Flash Payment System}  \label{sec:Flash:lowCongestion}

The Low-Congestion Flash Payment System, specified by Alg. \ref{alg:LCPS},  is geared for a setting in which payments are infrequent.  To achieve linear communication complexity in the good case, its blocks must be ``small'', namely of a constant size independent of the number of agents.  By \emph{issue a block} we mean create a block and send it to all agents.

\mypara{Issuing a small block} An agent $p$ issues a new block $b$ either when a payment to $p$ becomes final or when it wishes to make a payment.  When a block $b'$ with a payment to $p$ becomes final, the block $b$ it issues has two inputs: $b'$ and the previous $p$-block, and one output self-payment with the sum of the two input payments.  Assuming, inductively, that the previous $p$-block has a self-payment with the total balance of $p$, then the new self-payment in $b$ is the updated total balance of $p$.

When $p$ wishes to make a payment to an agent $q$, it issues a block $b$ with one input, the previous $p$-block, which includes $p$'s balance as a self-payment, and two output payments, one to $q$ and the remaining balance of $p$ (even if 0) as a self-payment.  

In both cases, the issued block $b$ includes one or two pointers to the roots of $[b] \setminus \{b\}$, which may or may not overlap with the input pointers specified above, as follows.
If the most recent block $b'$ received by $p$ is non-colliding, namely the local blocklace of $p$ has a single root $b'$, then $b$ includes a pointer to $b'$.  Otherwise, $b$ includes pointers to two roots of $p$'s local blocklace, one of which observes the oldest block (in order of receipt in the blocklace) still unobserved by any $p$-block.   Thus, in total, a block may have at most two output payments and four pointers -- two input pointers and two pointers to the blocklace roots.

In the good case, all agents behave correctly and  every issued block is received by all agents before any agent issues its next block.  In such a case there are no conflicts (equivocations or collisions) and  the protocol produces a linear chain of transaction blocks, each of a constant size.  A block is final when subsequent blocks issued by a supermajority of the agents approve it.
See Figure \ref{figure:goodcase}.A.
\begin{tcolorbox}[colback=gray!5!white,colframe=black!75!black]
\textbf{The Low-Congestion Flash Payment System:} 
\begin{enumerate}[leftmargin=*]
    \item \textbf{Transact \& Disseminate}: Upon a payment to $p$ becoming final or upon having a new payment to perform, issue a (small) block, and send to every agent $q$ not known to be faulty
    all blocks observed by the previously issued $p$-block but not by any received $q$-block. 
    \item \textbf{Receive}:  Upon receipt of a correct block pointing to blocks already in the blocklace, add it to the blocklace. 
\end{enumerate}
\end{tcolorbox}
Note that as the network is reliable, sending is realized so that each block is sent at most once to each destination.

\begin{figure}[tp]
\begin{algorithm}[H]
	\caption{\textbf{: Low-Congestion Flash Payment System}\\ Code for agent $p$, including Algorithm \ref{alg:blocklace}}	\label{alg:LCPS} 
	\begin{algorithmic}[1] \scalefont{0.93}
	  \alglinenoNew{counter}
   
\vspace{0.5em}
	\Upon{$\textit{transaction} \ne \bot $}   \Comment{1. \textbf{Transact \& Disseminate}} 
       \State   $\textit{issue}(\textit{new\_block}(\textit{transaction},\textit{`small'}))$
       \For{$\forall q \in P, b'\in [\textit{last\_block}(p)] : \lnot\textit{agentObserves}(q,b')$}
	    \State \textbf{send}  $b'$ to  $q$  \label{alg:lt-CordialDissemination}\Comment{\textbf{Cordial Dissemination}}
	    \EndFor   
        \State $\textit{transaction} \gets \bot $
	 \EndUpon

\vspace{0.5em}
	\Upon{\textbf{receive} $b : \textit{correct}(b) \wedge  [b]\subseteq \{b\}  \cup \textit{blocklace}  $}   \label{alg:BDA-receive}     
        \State $\textit{blocklace}\gets \textit{blocklace} \cup \{b\}$  \label{alg:BDA-receive} \Comment{2. \textbf{Receive}}	 
	 \EndUpon

\vspace{0.5em}
	\Upon{$\textit{utxo}(b,x) \in \textit{blocklace}$}   \label{alg:utxo}     
        \State $\textit{transaction}\gets (b,(p,x))$ \Comment{Add received unspent payment to balance} 
	 \EndUpon

		\alglinenoPush{counter}
	\end{algorithmic}

\end{algorithm}
\vspace{-2em}
\end{figure}

We make the following safety and liveness claims about the protocol, at the block level and at the transaction level.

\begin{proposition}[Safety of Low-Congestion \sys]\label{proposition:LC-safety}
If two blocks constitute a doublespend then it cannot be that both become final during a run of the protocol.
\end{proposition}
To prove this proposition, we first note that finality is final:
\begin{observation}\label{observation:finality}
Finality is monotonic wrt $\supset$.
\end{observation}
\begin{proof}
Let $B\subset B'$ be closed blocklaces and $b, b'\in B$, where $b'$ approves $b$ in $B$.  Then $b'$ also approves $b$ in $B'$. The reason is that since $B$ is closed, the set of blocks observed by $b'\in B$ is identical to the set of blocks $b'$ observes in $B'$, and hence if $b'$ observes $b$ but not observe a block equivocating with $b$ in $B$, the same holds also for $b'$ in $B'$.
\end{proof}
Secondly, we note that an agent cannot approve an equivocation without
being an equivocator itself (Fig. \ref{figure:approval}.A):
\begin{observation}[Only an Equivocator can Approve an Equivocation] \label{observation:approve-eq}
If agent $p \in P$ approves an equivocation in a blocklace $B$, then $p$ is an equivocator in $B$.
\end{observation}
\begin{proof}
By way of contradiction, assume that $p$ approves an equivocation $b_1, b_2 \in B$ via two blocks, $b_3$ and $b_4$ in $B$, respectively, that do not constitute an equivocation.  So w.l.o.g.\  assume that $b_3$ observes $b_4$.
(See Figure \ref{figure:approval}).  However, since $b_4$ observes $b_2$, then $b_3$ also observes $b_2$, and hence does not approve $b_1$.  A contradiction.
\end{proof}
An implication of this observation is that:
\begin{lemma}[No Supermajority Approval for Equivocation]\label{lemma:no-supermajority}
If  $b_1, b_2$ form an equivocation in blocklace $B$, then not both $b_1, b_2$ have  supermajority approval in $B$.
\end{lemma}
\begin{proof}
Assume that there are two supermajorities  $Q_1, Q_2 \subseteq  P$, where $Q_1$ approves $b_1$ and
$Q_2$ approves $b_2$.  A counting argument shows that two supermajorities must
have in common at least one correct agent. Let $p \in Q_1 \cap Q_2$
be such a correct agent.  Since $p\in Q_1$ it approves $b_1$ and since $p\in Q_2$ it approves $b_2$ by construction. Observation \ref{observation:approve-eq} shows that an agent that approves an equivocation must be a
an equivocator, hence $p$ is an equivocator.  A contradiction.
\end{proof}

We are now ready to prove safety:
\begin{proof}[Proof of Proposition \ref{proposition:LC-safety}]
By way of contradiction, assume that $b_1$ is final in some blocklace $B_1$ constructed during the execution, and $b_2$ is final in some other blocklace  $B_2$ constructed during the execution (it does not matter for the proof if these are global states, namely the union of the blocklaces of all agents at some point in the computation, or local blocklaces of specific agents).  By Observation \ref{observation:finality}, $b_1$ and $b_2$ are final in $B:= B_1 \cup B_2$, namely both equivocating blocks $b_1, b_2$ have supermajority approval in $B$. Lemma \ref{lemma:no-supermajority} claims this cannot be the case.  A contradiction.
\end{proof}

For liveness, we make the standard \textbf{fairness assumption} that a transition that is enabled infinitely often is eventually taken and make the \textbf{agent liveness assumption} that every correct agent eventually has a payment to make.  While we do not present a formal model of computation, we assume that a computation consists of a sequence of configurations, each specifying the local state of each agent, and that it proceeds via state transitions by the agents as specified by the protocol.

\begin{proposition}[Liveness of Low-Congestion \sys]\label{proposition:LC-liveness}
Any block by a correct agent will eventually become final, and every agent will eventually know that.
\end{proposition}

To prove the proposition, we first claim:
\begin{lemma}[Liveness of Cordial Dissemination]\label{lemma:liveness-dissemination}
A correct block known to a correct agent will eventually be known to every correct agent. 
\end{lemma}
\begin{proof}
Assume a correct block $b$ is known to $p$ and not known to $q$, both correct, at some configuration of the computation.
By the agent liveness and fairness assumptions, $p$ issues an unbounded number of blocks, and by construction each $p$-block observes the oldest block not observed by previous $p$-blocks.  Hence, eventually, $p$ will issue a $p$-block that observes $b$.

Then the next time $p$ issues a block, there are two cases:  Either $p$ has a $q$-block that observes $b$, in which case we are done.  Else, the condition for the Cordial Dissemination clause (Alg. \ref{alg:LCPS}, \cref{alg:lt-CordialDissemination}), 
that the previous $p$-block observes $b$ but no $q$-block received by $p$ observes $b$,  holds, resulting by fairness in $p$ eventually sending  $b$ to $q$.
Then  $q$ will eventually receive $b$ and accept it since it is correct by assumption.
\end{proof}

With this, we can prove the proposition:
\begin{proof}[Proof of Proposition \ref{proposition:LC-liveness}]
Let $b$ be a correct block issued by a correct agent $p$. By Lemma \ref{lemma:liveness-dissemination}, all correct agents will eventually know $b$.  By agent liveness and fairness and by the same argument as in the proof of the Lemma, every correct agent will eventually issue a block that observes $b$.  Since $p$ is correct, the blocks that observe $b$ do not observe a block equivocating with $b$, and therefore approve $b$, at which point $b$ is final by the set of blocks $B$ produced by all correct agents. 

Let $p$ be a correct agent.  Since every block in $B$ is produced by a correct agent, then according to Lemma \ref{lemma:liveness-dissemination}, $p$ will eventually know every block in $B$, at which point its local blocklace will be some superset of $B$.  Since finality is monotonic wrt to the superset relation (Ob. \ref{observation:finality}), then from that point on $p$ knows that $b$ is final.
\end{proof}

\mypara{Complexity and Latency}  We first consider the \textbf{good case}. Regarding communication complexity, each block is sent once to all agents (the Cordial Dissemination clause,  Alg. \ref{alg:LCPS}, \cref{alg:lt-CordialDissemination}, is not activated), 
and as the size of each block is constant,  the communication complexity is linear in the number of blocks. Ignoring self-payments, each output payment to $p$ requires two blocks, one that carries the payment to $p$ and a $p$-block that uses it as input to update $p$'s balance.  Hence the communication complexity per payment is $O(n)$.

While the protocol has no notion of rounds, it is useful to define it for latency analysis.  A \emph{round} is a fragment of the execution of the protocol in which each correct agent issues a block that is received by all correct agents.
Regarding latency, as a block is final when approved by a supermajority of the agents,  a block by a correct agent  becomes final following a single round.

Next, we consider the \textbf{worst case}. We note that in this low-congestion application, the practical relevance of the worst-case is primarily when the network is effectively down (highly asynchronous).  Malicious Byzantine activity is cared for, of course, but may be less relevant:  First, the only impact of malicious collisions is slowing the operation of the protocol, with no particular gain to colliders. Second, in the envisioned small-scale applications, Byzantines could be identified and excommunicated using out-of-band means.  Third, the protocol can be enhanced with methods for probabilistic identification  of Byzantine agents\cite{huang2023byzantine}, and then excommunicate intentional colliders within the protocol.

Regarding communication complexity, each block is sent once by all agents
to all agents by the cordial dissemination clause, resulting in communication complexity quadratic in the number of blocks, and hence the communication complexity per payment is $O(n^2)$.

Regarding latency, as in each round each correct agent observes an old block of a new agent $q$ (and all $q$-blocks received, if $q$ is correct),  then after a wave of $n$ rounds all correct agents would observe (and thus approve) all the blocks of all correct agents received prior to that wave.  In particular, a block by a correct agent, received by all correct agents, would become final following such an $n$-rounds wave, in the worst case.

\mypara{Urgent Payments} 
As block finalization in the Low-Congestion Flash Payment System may take a long time (a whole round of payments) even in the good case, a notion of urgency can be added, allowing a block with an urgent payment to become final in the next communication round, but at a communication cost of $O(n^2)$ (See Fig. \ref{figure:goodcase}.B). Note that, if agent $p$ needs urgent finalization of a block $b$ issued by another agent $q$ (for example because $b$ includes a payment to $p$), then $p$ can issue an urgent block $b'$ observing $b$, and the finalization of $b'$ will either also finalize $b$ or else expose a $q$-block equivocating with $b$, voiding the transaction in $b$.  If all payments are urgent, the communication complexity of the protocol is $O(n^2)$ and latency is one round, commensurate with the state-of-the-art.  The required addition to the Receive rule of the Low-Congestion Payment System is \underline{underlined} below.
\begin{tcolorbox}[colback=gray!5!white,colframe=black!75!black]
\textbf{An Addition for Urgent Blocks:} 
\begin{enumerate}[leftmargin=*]\setcounter{enumi}{1}
    \item \textbf{Receive \& Ack}:   Upon receipt of a correct block pointing to blocks already in the blocklace, add it to the blocklace.\\  \underline{If the block is urgent then issue an ack block}.
\end{enumerate}
\end{tcolorbox}
Urgent blocks are incorporated in Alg. \ref{alg:UB}. We note that urgent blocks do not affect the safety or liveness of the protocol.  A mechanism may be added by which issuing urgent blocks requires compensating the other agents for their extra work (issuing ack blocks).  Alternatively, a bound can be added on the number of urgent blocks issued by an agent as a function of the urgent blocks issued by other agents, to prevent Byzantine agents from abusing this mechanism.

\begin{figure}[tp]
\begin{algorithm}[H]
	\caption{\textbf{: Urgent Blocks added to Algorithm \ref{alg:LCPS}}\\ Code for agent $p$}	\label{alg:UB}
	\begin{algorithmic}[1] \scalefont{0.93}
	\alglinenoPop{counter}

\vspace{0.5em}
	\Upon{\textbf{receive} $b : \textit{correct}(b) \wedge  [b]\subseteq \{b\}  \cup \textit{blocklace}  $}   \label{alg:BDA-receive}     
        \State $\textit{blocklace}\gets \textit{blocklace} \cup \{b\}$  \label{alg:BDA-receive} \Comment{2. \textbf{Receive  \& Ack}}	
        \If{$\textit{urgent}(b.\textit{payments})$}  \Comment{Payments marked as urgent}
        \State   $\textit{issue}(\textit{new\_block}((b,\emptyset)),\textit{`small'}))$ \Comment{Acknowledge receipt of $b$}
        \EndIf
	 \EndUpon

		\alglinenoPush{counter}
	\end{algorithmic}

\end{algorithm}
\vspace{-2em}
\end{figure}

\subsection{The High-Congestion Flash Payment System}  \label{sec:Flash:highCongestion}

The High-Congestion Flash Payment System, specified by Alg. \ref{alg:HTPS},  is geared for a setting in which payment frequency is significantly higher than the block production rate.  In this case, to achieve linear communication complexity, its blocks must be ``large'', namely contain payments linear in the number of agents.  
This is achieved by the agent $p$ issuing a new block $b$  upon receipt of blocks from a supermajority of the agents, pointing to all the new blocks, regardless of whether and how many transactions it has.  The expectation is that by the time it receives a supermajority of new blocks it has already $O(n)$ payments to make.  Note that the new blocks would contain at most $O(n)$ new input payments.
Thus, in total, a block is expected to have $O(n)$ payments and to have size $O(n)$.

The protocol includes a parameter $\Delta\ge 0$ which is relevant for performance:
While the protocol is always correct in asynchrony, in the good case, i.e., when the network is synchronous, messages among correct agents arrive within $\Delta$, and $t \in O(1)$, the protocol incurs an $O(n)$ communication complexity. 

In the theoretical model of pure synchrony with a network delay $\Delta_0$, the wait parameter $\Delta$ can be adaptively chosen in each round to some $\epsilon \ge 0$ that completes the time from the previous round to $\Delta_0$, in which case the protocol would work at full network speed.

In practical systems, $\Delta$ can be tuned based on actual network performance. For example, assuming normal distribution of message arrival times, $\Delta$ can be set to 2-3 standard deviations ($\sigma$), thus allowing in expectation $>97\%$-$99.9\%$ of the blocks to join each round in the good case, not just the first supermajority that arrives, at the cost of a small delay ($2\sigma$-$3\sigma$)  per round. In either case, the value of $\Delta$ has no impact on the correctness of the protocol.

Protocol rounds are organized according to block height.  The \emph{height} of a 
correct $p$-block $b$ is  1 if $b$ is initial, else it is $1+$ the maximal height of any block $b$ observes. 


\begin{tcolorbox}[colback=gray!5!white,colframe=black!75!black]
\textbf{The High-Congestion Flash Payment System:}\\
Set $k:=1$ and issue an initial block.
\begin{enumerate}[leftmargin=*]
    \item \textbf{Transact}:  Upon the blocklace having a supermajority of blocks of height $k$, wait $\Delta$ and issue a (large) block of height $k+1$.
    \item \textbf{Receive \& Disseminate}:  Upon receipt of a correct $q$-block $b$ of height $k'$ pointing to blocks already in the blocklace, add it to the blocklace and send to $q$ all blocks of height $< k'$  not observed by $b$. 
\end{enumerate}
\end{tcolorbox}
As in the Low-Congestion variant, sending is realized so that each block is sent at most once to each destination.

The disadvantage of the high-congestion variant is that, unlike the low-congestion variant, it requires all agents to issue, receive, store, and process blocks at full network speed, even if they have no payments to contribute.  Furthermore,  if the payment rate is too low, resulting in batching less than $n$ payments per block, then linear communication complexity per payment it not attained.

\begin{figure}[tp]
\begin{algorithm}[H]
	\caption{\textbf{: High-Congestion Flash Payment System}\\ Code for agent $p$, including Algorithm \ref{alg:blocklace}}	\label{alg:HTPS}
	\begin{algorithmic}[1] \scalefont{0.93}
	\alglinenoPop{counter} 
	\Statex \textbf{Local variables:}
      
        \StateX  $\Delta$: \Comment{Wait time, see discussion in text}
   
\vspace{0.5em}

        \State   $\textit{issue}(\textit{new\_block}((\emptyset,(p,x_p))))$ \Comment{Initial block and round, $x_p$ is the initial balance of $p$}

\vspace{0.5em}
	\Upon{$\textit{cordial}$}   \Comment{1. \textbf{Transact}} 
        \State \textbf{wait} $\Delta$
        \State   $\textit{issue}(\textit{new\_block}(\textit{transaction},\textit{`large'}))$   
	 \EndUpon

\vspace{0.5em}
	\Upon{\textbf{receive} $b : \textit{correct}(b) \wedge  [b]\subseteq \{b\}  \cup \textit{blocklace}  $}   \label{alg:BDA-receive}     
        \State $\textit{blocklace}\gets \textit{blocklace} \cup \{b\}$  \label{alg:BDA-receive} \Comment{2. \textbf{Receive  \& Disseminate}}	
        \For{ $\forall b'\in [\textit{last\_block}(p)] : \lnot\textit{agentObserves}(b.\textit{creator},b')$}
	    \State \textbf{send}  $b'$ to  $b.\textit{creator}$  \label{alg:send-package}\Comment{\textbf{Cordial Dissemination}}
	    \EndFor    
	 \EndUpon

        \vspace{0.5em}
        \Procedure{\textit{cordial}}{}
        \State \Return $\{b.\textit{creator} : b \in \textit{blocklace}\setminus [\textit{last\_block}(p)]\}$ is a supermajority
        \EndProcedure

		\alglinenoPush{counter}
	\end{algorithmic}

\end{algorithm}
\vspace{-2em}
\end{figure}

\begin{proposition}[Safety of High-Congestion \sys]\label{proposition:HC-safety}
If two blocks constitute a doublespend then not both become final during a run of the protocol.
\end{proposition}
\begin{proof}
The proof is the same as the proof of Proposition \ref{proposition:LC-safety}. 
\end{proof}

\begin{proposition}[Liveness of High-Congestion Payment System]\label{proposition:HC-liveness}
Any block by a correct agent will eventually become final, and every correct agent will eventually know that.
\end{proposition}
\begin{proof}
In each round, every correct agent awaits a supermajority of new blocks and then issues a block that observes all correct received blocks, so liveness as claimed by the proposition is ensured as long as the protocol does not deadlock.
Hence, we have to show that for each round completed by an agent $p$, a supermajority of the blocks of that round will be received by $p$, enabling it to proceed to the next round.  We prove this inductively for the height $k$.

If $k=1$ then all correct agents issue an initial block, and since the network is reliable, eventually $p$ will receive a supermajority of the initial blocks of height 1 and issue a height 2 block.
Assume this holds for height $k$, meaning that all correct agents have issued blocks of height $k$.  Then eventually $p$ will receive all blocks by correct agents of height $k$.  

However, these blocks may depend on blocks of height $<k$ not known to $p$,  hence preventing $p$ from accepting such blocks.  Since the block of height $k$ issued by $p$ discloses to all correct agents the blocks $p$ knows of height $<k$, its receipt will trigger Cordial Dissemination to $p$ of these blocks upon receipt.  Hence, $p$ will eventually receive  from each correct agent $q$ all such blocks of height $<k$ that the $q$-block $b$ of height $k$ depend upon, allowing $p$ to receive $b$.
Once $p$ receives same from a supermajority of the agents, it can issue the block of height $k+1$.
\end{proof}

\mypara{Complexity and Latency}  
The analysis assumes that the frequency of payments is much higher than the pace of rounds of block creation and dissemination, so that for each round each correct agent accumulates a linear number of payments to be made per block.
Also, here we measure rounds by block height: The round $k$ is completed when a supermajority of agents issue blocks of depth $k$ that are received by a supermajority of the agents.

We first consider the \textbf{good case}.  Regarding communication complexity, each block is sent once to all agents (the Cordial Dissemination clause is not activated), and as the size of each block is linear,  the communication complexity is quadratic in the number of blocks. However, as each block has a linear number of payments,  the communication complexity per payment is $O(n)$.

Regarding latency, as a block is final when approved by a supermajority of the agents, a block by a correct agent  becomes final following a single round.

Next we  consider the \textbf{worst case}, in which we still assume that for each round each correct agent accumulates a linear number of payments to be made per block.
Regarding communication complexity, each block is sent once by all agents
to all agents by the cordial dissemination clause, resulting in communication complexity cubic in the number of blocks, and the communication complexity per payment is $O(n^2)$.

Regarding latency, a block by a correct agent  becomes final following a single round as in the good case.
However,  as a $q$-block $b$ of round $k$ cannot be received  by a correct agent $p$ if $b$ observes blocks not known to $p$, a round as defined may take one more communication round to complete:  Once $q$ receives the $p$ block of round $k$ it would know which round-$k-1$ blocks $p$ possibly does not know yet, and send them to $p$.

\section{Related Work} \label{sec:relatedWork}

A few works~\cite{pedone2002handling,gupta2016non} observed that a deterministic financial payment system can be implemented in asynchrony. 
The first formal definition of an account-based payment system as a shared memory object was done by Guerraoui et al.~\cite{guerraoui2019consensus}.
They showed that the consensus number~\cite{herlihy1991wait} of a payment system is one, i.e., it is a weaker problem than consensus.

Guerraoui et al. also provided a concrete algorithm implementing a payment system in the message-passing asynchronous model, circumventing the seminal FLP result~\cite{fischer1985impossibility}.
Their algorithm is based on reliable broadcast~\cite{bracha1987asynchronous} that preserves the order of messages originating from the same agent~\cite{malkhi1997high}, i.e., if an agent sends message $m_1$ before $m_2$, then $m_1$ arrives at all honest agents before $m_2$.
The communication complexity of this algorithm is $O(n^2)$ per payment in the good and worst case. This complexity matches the worst case lower bound which was proven in~\cite{naor2022payment}.

Another definition of a payment system is due to Auvolat et al.~\cite{auvolat2020money}.
They define a weaker specification for a payment system than the one in~\cite{guerraoui2019consensus}, which allows them to provide an implementation using a reliable broadcast algorithm that guarantees
FIFO order between every two agents.
The communication complexity of their algorithm is also $O(n^2)$.

Astro~\cite{collins2020online} is a concrete implementation of a payment system based on the definition in~\cite{guerraoui2019consensus}.
It uses reliable broadcast and has two modes of operation: the first mode simply uses Bracha's reliable broadest algorithm~\cite{bracha1987asynchronous} which has a $O(n^2)$ communication complexity. 
In the second mode, when an agent wishes to make a payment, it simply sends a message to all other agents, thus not using reliable broadcast, and so forgoing some of the reliable broadcast properties.
Instead, to prevent double-spend, when the agent sends a payment message, it also includes $2f+1$ ack messages previously sent to it justifying that the payment is valid, making each payment message $O(n)$ in size. 
Hence, the overall communication complexity of this protocol in the good and worst case is also $O(n^2)$, as are the previously mentioned protocols.

As mentioned in the introduction, to the best of our knowledge, \sys is the first payment system that exhibits a linear communication complexity per payment in the good case, while still incurring the lower bound quadratic communication complexity in the worst case.

Besides improving the message complexity of asynchronous payment networks, there are a few works on asynchronous payment channels~\cite{naor2022payment,ersoy2022syncpcn} to scale a payment system performance. 
A payment channel acts as a second layer on top of a payment system between a payer and a payee.
To this end, the payer ``locks'' a deposit that funds the payment channel using the payment system (incurring quadratic communication complexity), and then it can make multiple payments on the channel which requires the payer to send a single message to the payee.
To end the channel use, the underlying payment system is used again to move the locked deposit to reflect the last state of the channel.
Since the intermediate payments on the channel do not use the payment system, it removes those payments (and the associated messages) from it.

Using a DAG-like structure to represent dependencies between blocks has been previously introduced, for example, in Byteball~\cite{churyumov2016byteball}, Vegvisir~\cite{karlsson2018vegvisir}, Corda~\cite{hearn2016corda}, the GHOST protocol~\cite{sompolinsky2015secure}, DAG Knight~\cite{sompolinsky2022dag}, DAG-Rider~\cite{keidar2021need}, Aleph~\cite{gkagol2018aleph}, Narwhal and Tusk~\cite{danezis2021narwhal}, Bullshark~\cite{danezis2021narwhal}, and Cordial Miners~\cite{keidar2022cordial}.
Albeit these protocols use a distributed structure to represent dependencies or partial order between blocks, in the end they still employ consensus to totally order the blocks in the DAG into a single linear blockchain.

\section{Conclusion} \label{sec:conclusion}

We presented \sys, a payment system that, for the first time, demonstrates linear communication complexity when the number of Byzantine faults is low, and quadratic complexity when it is high, matching the lower bound in the worst case.
\sys has two modes of operation for two distinct scenarios: for low congestion where payments are issued infrequently, and for congested networks where multiple agents issue multiple payments concurrently.

\begin{acks}
Oded Naor is grateful to the Azrieli Foundation for the award of an Azrieli Fellowship, and to the Technion Hiroshi Fujiwara Cyber-Security Research Center for providing a research grant.
Ehud Shapiro is the Incumbent of The Harry Weinrebe Professorial Chair of Computer Science and Biology at the Weizmann Institute.  We thank Nimrod Talmon for his comments on an earlier version of the manuscript and Idit Keidar for pointing us to related work.
\end{acks}

\bibliographystyle{ACM-Reference-Format}
\bibliography{bib}

\newpage
\appendix

\section{Blocklace Utilities}\label{appendix:blocklace}

\begin{figure}[th!]
\begin{algorithm}[H]
    \caption{\textbf{: Blocklace Utilities} \\
    Code for agent $p $}
    \label{alg:blocklace}
    \small
  
    \alglinenoPop{counter} 

    \begin{algorithmic}[1] \scalefont{0.9}
        \Statex \textbf{Local variables:}
      
        \StateX struct $\textit{h}$: \Comment{The structure of a hash pointer $h$ in a block}
        \StateXX $h.\textit{hash}$ -- the value of the hash pointer
        \StateXX $h.\textit{creator}$ -- the signatory of the hash pointer

        \StateX struct $\textit{transaction}$: \Comment{Updated externally. A transaction is a pair (Inputs, Output payments)}
        \StateXX $\textit{transaction}.\textit{inputs}$ -- pointers marked as inputs 
        \StateXX $\textit{transaction}.\textit{payments}$ -- output payments,  pairs $(\textit{agent},\textit{amount})$

        \StateX struct $\textit{block } b$: \Comment{The structure of a block $b$ in a blocklace}
        \StateXX $b.h$ -- hash value of the rest of the block, signed by its creator $p$ 
        \StateXX $b.\textit{payments}$ -- output payments 
        \StateXX $b.\textit{pointers}$ -- a possibly-empty set of signed hash pointers to other blocks, some marked as inputs

	  \StateX \textit{blocklace} $\gets \{\}$  \Comment{The local blocklace of agent $p$}
   
		\Procedure{$\textit{new\_block}$}{\textit{transaction,\textit{type}}}  
		\State \textbf{new} $b$  \Comment{Allocate a new block structure}
		
        \State $b.\textit{payments} \gets \textit{transaction}.\textit{payments}$ 
        \State $b.h \gets \textit{hash}((b.\textit{pointers},b.\textit{transaction}))$ signed by $p$
        \If{\textit{type}=\textit{`large'}}
        \State $b.\textit{pointers} \gets \{h :  h.\textit{hash} := b.\textit{hash} \wedge 
            h.\textit{creator} := b.\textit{creator} \wedge b \in \textit{roots}\} \cup \textit{transaction}.\textit{input}
            $ \label{alg:roots} 
        \EndIf
         \If{\textit{type}=\textit{`small'}}
        \State $b.\textit{pointers} \gets \{h :  b:= \textit{last\_block}(p) \wedge h.\textit{hash} := b.\textit{hash} \} \cup \textit{transaction}.\textit{input}$
             \label{alg:roots} 
        \EndIf
        \State \Return $b$
		\EndProcedure

        \Procedure{$\textit{observes}(b,b')$}{} 
        \State \Return $\exists b_1,b_2,\ldots,b_k \in \textit{blocklace}$, $k\ge 1$,  s.t.\
        $b_1 = b $, $b_k = b'$ and $\forall i \in [k-1] \exists h \in b_i.\textit{pointers} \colon h.\textit{hash}= \textit{hash}(b_{i+1})$
        \EndProcedure

        \Procedure{$\textit{roots}$}{} 
        \State \Return $\{b : b \in \textit{blocklace} \wedge \lnot\exists b'\in \textit{blocklace}$ s.t. $b'\succ b\}$
        \EndProcedure

	    \Procedure{\textit{equivocation}}{$b_1,b_2$}: \label{alg:SMR:da}
            \Return 
            $b_1 \not\succ b_2 \wedge  b_1 \not\succ b_2 \wedge b_1.\textit{creator} = b_2.\textit{creator} 
           $ \Comment{Fig. \ref{figure:approval}.A} 
        \EndProcedure

	    \Procedure{\textit{equivocator}}{$q,b$}: \label{alg:SMR:dactor}  \Comment{Fig. \ref{figure:approval}.A}
            \State \Return 
            $\exists b_1,b_2 \in [b] \wedge
            b_1.\textit{creator} = b_2.\textit{creator} = q \wedge
          \textit{equivocation}(b_1,b_2)
           $
        \EndProcedure

		\Procedure{\textit{approves}}{$b,b_1$}: \label{alg:SMR:approves}
            \Return $b_1 \in [b]  \wedge \forall b_2 \in [b] : \lnot$\textit{equivocation}$(b_1,b_2)$  \Comment{Fig. \ref{figure:approval}}
        \EndProcedure

		\Procedure{\textit{final}}{$B,b$}: \Comment{Fig. \ref{figure:approval}} \label{alg:ratifies} 
         \State \Return $\{b'.\textit{creator} :  b' \in \textit{blocklace} \wedge \textit{approves}(b',b)\}$ is a supermajority
            \EndProcedure

        \Procedure{$\textit{last\_block}$}{$p$} 
        \State \Return $b \in blocklace$ s.t. $b.\textit{creator}=p \wedge 
           ( \forall b'\in blocklace : b'.\textit{creator}=p \implies b'\not\succ b)$
            \EndProcedure

        \vspace{0.5em}
        \Procedure{\textit{issue}}{$b$}
        \For{all $q \in P$} \textbf{send}  $b$ to  $q$  \Comment{\textbf{send} is idempotent;  $b$ is sent to $q$ at most once}
        \EndFor
        \EndProcedure

        \vspace{0.5em}
        \Procedure{\textit{utxo}}{$b,x$}
        \State \Return  $b\in \textit{blocklace}$ includes a payment $(p,x)$ and no $p$-block in \textit{blocklace} points to $b$ and $\textit{final}(\textit{blocklace},b)$.
        \EndProcedure
        \alglinenoPush{counter}
 
    \end{algorithmic}
\end{algorithm}
\vspace{-2.5em}
\end{figure}


\end{document}